# Compositional gradients in sputtered Ti-Au alloys: Site-selective Au-decoration of anodic TiO$_2$ nanotubes towards enhanced photocatalytic H$_2$ evolution

*Seyedsina Hejazi,[a] Marco Altomare,[a] Shiva Mohajernia[a] and Patrik Schmuki[a,b],\**

S. Hejazi, Dr. M. Altomare, S. Mohajernia and Prof. Dr. P. Schmuki*

[a] Department of Materials Science and Engineering WW4-LKO University of Erlangen-Nuremberg Martensstrasse-7, Erlangen D-91058, Germany

[b] Chemistry Department, Faculty of Sciences, King Abdulaziz University, 80203, Jeddah, Saudi Arabia

\* Corresponding author. E-mail: schmuki@ww.uni-erlangen.de

**Abstract**

Au nanoparticles at the $TiO_2$ surface can enhance the photocatalytic $H_2$ generation performances owing to their electron transfer co-catalytic ability. Key to maximize the co-catalytic effect is a fine control over Au nanoparticle size and placement on the photocatalyst, in relation to parameters such as the $TiO_2$ morphology, illumination wavelength and pathway, and light penetration depth in the photocatalyst. Here we present an approach for site-selective intrinsic-decoration of anodic $TiO_2$ nanotubes (TNs) with Au nanoparticles: we produce, by Ti and Au co-sputtering, Ti-Au alloy layers that feature compositional gradients across their thickness; these layers, when anodized under self-ordering electrochemical conditions, can form Au-decorated TNs where the Au nanoparticle density and placement vary according to the Au concentration profile in the metal alloy substrates. Our results suggest that, the Au co-catalyst placement strongly affects the photocatalytic $H_2$ evolution performance of the TNs layers. We demonstrate that, when growing Au-decorated TNs, the use of Ti-Au substrates with a suitable Au compositional gradient can lead to higher $H_2$ evolution rates compared to TNs classically grown with a homogenous co-catalyst decoration. As a side effect, a proper placement of the co-catalyst nanoparticles allows for reducing the amount of noble metal without dumping the $H_2$ evolution activity.

Ever since the first report of Fujishima and Honda in 1972 on photo-electrochemical water splitting,[1] titanium dioxide ($TiO_2$) has obtained huge attention due to its semiconducting properties that make it suitable for a wide range of photo-electrochemical and photocatalytic applications, as well as to fabricate sensors and electronic devices, among others.[2–4]

Photocatalysis is based on the interaction of light with a semiconductor immersed in a suitable reaction medium. Photons of adequate energy (higher than the semiconductor band gap) can excite electrons from the valence band (VB) of the semiconductor to its conduction band (CB), creating electron-hole pairs ($e^-$-$h^+$). Once separated, holes and electrons can reach the surface of the semiconductor and react with redox species in the environment. One feature of $TiO_2$ is that the (anatase) $TiO_2$ CB minima lies above the electrochemical potential of $H_2O$ reduction;[5] i.e. $TiO_2$ CB electrons generated upon UV light illumination are sufficiently energetic to reduce water to $H_2$ gas.

Among various $TiO_2$ structures investigated in photocatalysis, nanostructures such as nanotubes have attracted large interest in the last decades owing to their large surface area, specific charge separation features as well as due to their easy and versatile fabrication method. For example, ordered arrays of TNs can be grown by anodic oxidation of a Ti metal substrate in a fluoride-containing electrolyte. This approach allows for a high control over the nanotube geometry (length, diameter, wall thickness).[6,7] Nevertheless, pristine $TiO_2$ is not an efficient photocatalyst for $H_2$ generation. Main reasons are the sluggish kinetics of electron transfer and the rapid recombination rate of photogenerated charge carriers in $TiO_2$.[8] Therefore, a charge transfer co-catalyst is required, and most common co-catalysts are based on noble metals such as Au, Pd or Pt, which are typically deposited in the form of nanoparticles (NPs) on the $TiO_2$ surface. These noble metal NPs can act as electron transfer mediators and $H_2$ recombination catalyst.[9,10]

Conventional methods developed for decorating noble metal NPS on a given photocatalyst surface are chemical deposition,[11] sputtering,[12–14] and photo-deposition[15–17] among others. More recently, Lee et al. demonstrated that the anodization of Au-Ti cast alloy substrates under self-organizing electrochemical conditions can lead to Au nanoparticle-decoration of $TiO_2$ NTs.[18] More generally, when Ti-X alloys are used for self-organizing anodization, and X is e.g. a "valve" metal such as Nb and Ta[19–21], the alloying element X is simultaneously oxidized during the electrochemical treatment and is incorporated into the $TiO_2$ nanotube structure (as a secondary oxide, or as dopant in the $TiO_2$ lattice). The situation is different when X is a noble metals such as Au: the noble metal is not oxidized during the anodization process but forms, via self-diffusion and agglomeration, noble metal clusters at the TNs surface. Previous work was carried out using Ti-X cast alloys that, owing to their intrinsic homogeneous composition, were found to lead to a homogeneous surface decoration of the $TiO_2$ tubes – e.g., in the case of anodic layers grown from 0.2 at.% Ti-Au cast alloys, the result is TNs that are homogeneously coated along their sidewalls with ~ 3.5 nm-sized Au nanoparticles.[18]

Recently we reported on "metastable" Ti-Au metal substrate produced by Ti and Au co-sputtering that feature Au concentrations that can exceed the solubility limit in cast Ti-Au alloys. We showed the use of such substrates for the anodic formation of $TiO_2$ NTs with a much higher density of Au NP loading than by using classic alloys [22]. Here we fabricate Ti-Au alloy layers by co-sputtering and by systematically varying the deposition rate of Au during the co-sputtering process. The resulting Ti-Au alloy layers feature tunable Au compositional gradients across their thickness. We demonstrate that, when these Ti-Au alloy substrates are anodized, the resulting TNs feature a site-selective Au NP decoration, that is, the density of Au NP decorations along the nanotube walls varies according to the Au composition profile in the Ti-Au alloy substrate. We control the Au decoration placement to enhance the Au-$TiO_2$ NT photocatalytic performance and can achieve much higher $H_2$ evolution rates compared to a classic homogeneous co-catalyst decoration of $TiO_2$ NTs with comparable Au loadings.

The Ti-Au gradient alloy layers were prepared by sweeping (e.g. increasing or decreasing) the Au deposition rate during the co-sputtering process, while the Ti deposition rate was kept constant. When the Au deposition rate is increased during the sputtering process, the resulting Ti-Au layer show a Au concentration profile that increases across its thickness from the bottom to the top (Fig. S1). Such layer is here labeled as "$G_{top}x$"; "$G$" stands for "*gradient*" alloy, and "*x*" is the overall Au content in the layer in at.%, determined by EDX analysis. On the contrary, if the Au deposition rate is gradually reduced during the co-sputtering process, the resulting Ti-Au layers have a Au concentration profile that decreases from the bottom to the top; this layer is labeled as "$G_b x$".

For comparison, Ti-Au homogenous alloys were fabricated by co-sputtering of Ti and Au metals at constant rates. Various Au concentrations in the sputtered alloy layers could be obtained by adjusting the Au and Ti relative deposition rate (i.e. deposition power – see Fig. S2), which leads to an adjustable Au content in the alloy in the 0-2.1 at.% range. The homogeneous alloys are labeled as "$Hx$", where "$H$" stands for "homogeneous" alloy and "*x*" is the overall Au content in the metal layer (in at.%).

To grow the Au-decorated TNs, 750 nm-thick Ti-Au layers were anodized in an ethylene glycol electrolyte, with 0.15 M HF and 3 wt.% $H_2O$, at 60 V. During the electrochemical anodization, the Ti metal acts as back contact meanwhile undergoing conversion at the outermost surface into arrays of nanotubular structures. The anodization time was 2 min. Typically, after such anodization time, the metal layers were fully converted into 2 µm-thick vertically-aligned self-organized TNs layers (Fig. S3). Current density (*j*) – time profile are a most useful tool to follow the anodization process (see Fig. S4a); when no Ti metal is left at the bottom of the tubular structures, $j$ decreases, and the anodization is interrupted. In this electrochemical configuration, the Si wafer substrate is merely a support for the formed TNs.

Figure 1a,c,e show the top and bottom structures of anodic TNs grown from homogenous Ti-Au sputtered layers with a Au content of 0.9 at.% (*H0.9*). In contrast to the homogeneity of the Au NP decoration of tubes grown from *H0.9* alloys, tubes grown from gradient Ti-Au substrates (e.g. $G_{top}0.6$) are site-selectively decorated, as shown in the SEM images in Figure 1b,d,f: in line with the Au concentration profile in the Ti-Au gradient substrate, the Au NPs form only at the top of the tubes, while the tube bottom remains pristine.

Aside from the differences in Au NP placement, TNs grown from homogeneous or gradient substrates appear otherwise identical; their length, wall thickness and inner diameter are 2 µm, 30 nm, and 50 nm, respectively (Fig. 1c). In any case, the expansion factor (ratio between the thickness of the formed oxide layer vs. that of the original metal layer) is approximately 2.7, which fits well to the data in the literature.[23] In this range of Au concentration, the incorporation of Au into the Ti metal layers, either in a homogeneous or in a gradient-like fashion, doesn't seem to influence the nanotube morphology.

We performed GDOES analysis to investigate the Au concentration profile across the NT layers – GDOES sputter profile data are compiled in Fig. 1g, h. In line with the SEM results, the concentration of Au in tubes grown from homogeneous Ti-Au alloys (*H0.9*) is constant across the full NT length. The NT oxide/Si wafer interface appears with the simultaneous gradual decrease of the O signal and appearance of Si (see Fig. S5). The overall Au content measured in the whole layer by EDX analysis is 0.6 at.%. On the contrary, the Au concentration profile for NTs grown from gradient substrates ($G_{top}0.6$) shows that Au is concentrated at the NT top, and the Au signal gradually fades off while approaching the tube bottom. Also in this case, the overall Au content in the whole layer is 0.6 at.% (EDX analysis).

XRD analysis of the sputtered Ti-Au metal layers and of the resulting TiO$_2$ NT arrays was performed to acquire information on their crystallographic phase composition. The XRD patterns of the homogenous (*H0.9*) and gradient (*G$_{top}$0.6*) sputtered alloys are shown in Fig. 2a, and are compared with that of a sputtered pure Ti substrate. The pure Ti metal layer show an hexagonal close packed (hcp) crystal α phase structure, and the position of the XRD reflections correspond well with those reported in the literature for metallic Ti (JCPDS card no. 44-1294). XRD patterns of homogenous (*H0.9*) and gradient (*G$_{top}$0.6*) alloys are virtually identical to each other and to the pattern of sputtered pure Ti metal layers; apparently for such Au concentrations, the crystal structure of the sputtered Ti-Au alloy layers is not affected by the incorporation of Au.

Current-time (*j-t*) characteristics were recorded during the anodization of sputtered pure Ti layers, as well as during the anodization of homogenous and gradient Ti-Au alloys (Fig. 2b). In line with the literature,[24] the *j*-t curve for anodization of pure Ti layers shows the typical profile observed for the anodic growth of TiO$_2$ NT layers: that is, after a steep increase of current density upon applying the anodic bias, a compact oxide layer is formed in the early stage of the anodization and *j* accordingly decreases; afterwards, irregular nanoscale pores form into the initial compact oxide and then an array of regular nanotubular structures starts forming. From this point on, the process reaches steady-state conditions, i.e. a constant *j* value, which indicates that an equilibrium is established between the oxide NT formation at the metal/oxide interface and the anodic oxide dissolution at the oxide/electrolyte interface.[24]

*J-t* curves of anodic layers grown from sputtered homogeneous Ti-Au layers (*H0.9*, 0.9 at.% Au) look virtually identical to that of sputtered pure titanium layers (Ti). A minor difference is that the steady state current seems to slightly increase in the presence of Au; we also found that the higher the Au content the higher the steady state anodic current; this can be seen when comparing the *j-t* profiles of tubes grown from metal layers with different Au content, e.g. pure Ti, *H0.1* and *H0.9* (Fig. S4b). The situation is clearly different for tubes grown from gradient alloys (*G$_{top}$0.6*): the *j-t* curve (Fig 2b) shows a clearly higher current density during the preliminary stage of the anodization process. The reason can be a more pronounced gas evolution due to a higher concentration of gold at the top of the Ti-Au layer. However, as the anodization front moves inward the Ti-Au substrate, the anodic current diminishes as a consequence of the gradually lower Au concertation.

After the electrochemical growth, the TNs were crystallized by annealing at 450°C (1h, in air). The XRD diffractograms of these structures are compiled in Fig. 2c. The reflections at 25.6° and 37.8° confirm the formation of $TiO_2$ anatase phase (JCPDS Card no. 21-1272).[25] Moreover, the XRD pattern of tubes grown from the gradient alloys (*G$_{top}$0.6*) show a reflection at 44.6° that can be attributed to Au cubic phase (JCPDS card no. 04-0784). Interestingly, this peak does not appear in the XRD pattern of NTs grown from homogenous alloys (*H0.9*); the reason can be the higher loading of gold clusters at the NT top (localized Au decoration) when growing tubes from the gradient alloy (*G$_{top}$0.6*); while the Au loading in homogeneously decorated NTs may be below the XRD detection limit.

In order to evaluate the photocatalytic behavior of the Au decorated TNs grown from different sputtered Ti-Au layers (homogeneous and gradient-like), their photocatalytic $H_2$ evolution rate ($r_{H2}$) from water-ethanol solution was measured under UV light irradiation (see Fig. 3a).

The data in Fig. 3a shows that pristine TNs, i.e. grown from pure sputtered Ti layers (Ti), deliver a negligible $H_2$ generation performance. On the contrary, the $H_2$ evolution rate is higher for all the Au-decorated structures. In general, this is ascribed to the formation of localized $Au/TiO_2$ Schottky junctions along the $TiO_2$ NT sidewalls; Au co-catalytic nanoparticles can act as electron-trap for $TiO_2$ CB electrons, thereby facilitating charge separation, and as electron transfer mediator by transferring electrons to the environment for H$_2$ evolution. This can limit electron-hole recombination in TiO$_2$, consequently leading to enhanced photocatalytic performances.[26,27]

However it is evident that the H$_2$ evolution activity is strongly influenced by the Au loading, which can be controlled by adjusting the Au content in the sputtered Ti-Au layers. The Au content in the sputtered Ti-Au metal substrates and the Au loading on the different tubes structures is reported in Fig. 2c,d . The results reveal a similar Au concentration compared to the sputtered Au-Ti layers.

In the case of homogeneous Ti-Au layers, it is found that an increase of the Au content in the alloy, i.e. Au concentration increasing from 0.1 to 0.9 at.%, leads to an increase of the Au loading on the NTs (Fig. 3a), which in turn causes a significant enhancement of the $r_{H2}$ (Fig. 3b). Au-TiO$_2$ NTs grown from *H0.9* lead to at $r_{H2}$ of 34 µL h$^{-1}$ cm$^{-2}$, which is more than 8 times higher the H$_2$ evolution of tubes grown from H0.1 (4.2 µL h$^{-1}$ cm$^{-2}$). The morphology of the TNs layer formed from *H0.1* substrates is shown in Fig. S6. Thus, the observed photocatalytic improvement seems to be ascribed to an increase of the Au decoration density on the NTs.[18,28] Nevertheless, in an attempt to further increase the Au NP density, we found that the structures grown from homogeneous alloys with a Au content of 2 at.% (*H2*) show a significantly lower $r_{H2}$ (2.1 µL h$^{-1}$ cm$^{-2}$). The reason for this is that these substrates, owing to their relatively high Au content, do not grow ordered NTs but only undefined Au-TiO$_2$ porous structures – see Fig. S7 and the SI for more details. In addition, these anodic layers are only 400 nm-thick while NT layers grown from Ti-Au substrates with Au content < 2 at.% are 2 µm-thick; thus, the poor photocatalytic activity of tube grown from H$_2$ substrates can be caused by a non-optimized light absorption (photon harvesting).

Noteworthy, top decorated NTs grown from top-gradient Ti-Au substrates (*G$_{top}$0.6*) are, in spite of the lower Au content in the metal substrate, as active as homogeneously decorated tubes grown from substrates with Au content of 0.9 at.%. Even more remarkable is that if the photocatalytic data are normalized vs. the Au loading in the Ti-Au substrates (Fig. 3c) the Au loading seems not to influence the photocatalytic activity in the case of NT structures grown from homogeneous decoration. However, the situation is different when the NTs are site selectively decorated: top decorated tubes ($G_{top}0.6$) are for comparable a Au loading (i.e. 0.63 at.%) more active not only than bottom decorated tubes ($G_{bottom}0.6$) but also than any other homogeneously decorated NT layers (additional SEM data for NTs grown from gradient alloys $G_b0.6$ are in Fig. S8). The reason for this can be related to the UV light penetration depth in the Au-TiO$_2$ NT layers [14] with respect to the Au NP distribution and to the electron diffusion length [14]. Yoo et al. reported that for noble metal-decorated TiO$_2$ NT layers, a most efficient photocatalytic configuration implies a direct illumination of the NT/co-catalyst/environment interface.[14,29–31] Thus, the co-catalytic effect is maximized when the Au NPs are placed at the NT top, i.e., the part of the TNs that is directly illuminated and where a relatively high density of charge carrier can therefore be generated and efficiently transferred to the environment for H$_2$ evolution. The situation is different for the tube bottom where, owing to light attenuation effects, the charge carrier density is anticipated to be lower. In other words, Au-TiO$_2$ NTs grown from top-gradient Ti-Au alloys allow for a most efficient use of the co-catalyst, i.e. by a site-selective intrinsic Au decoration of the highly-reactive NT top.

In summary, we showed that Ti-Au co-sputtering allows for the fabrication of TiAu alloy layers with controllable Au content. Key of our approach is that Au can be incorporated not only evenly throughout the sputtered layers, leading by anodization to homogeneously decorated Au-TiO$_2$ NTs, but also with desired composition profiles where the Au content gradually varies across the layer thickness (i.e. Au concentration gradients). We demonstrate that such Au concentration gradients lead to site-selectively Au-decorated TNs. Compared to TNs that carry a homogeneous Au decoration, a proper placement of the Au nanoparticles can maximize the Au co-catalytic effect, leading to a higher H$_2$ evolution rate per mass of used Au co-catalyst. More generally, the concept outlined in the present work may be extended to other

Ti alloys for the growth of a large palette of site-specifically functionalized TNs. While in the present study we prove that the metal decoration density can feature gradient along the nanotube walls, we anticipate that noble metal or metal oxide can also be embedded at specific depths within the TNs in the form of sharp localized junction or periodic compositional paradigms.

**Experimental Section**

Metastable Ti-Au layers were prepared by a SP-P-US-6M-32 Createch magnetron-sputtering device. Co-deposition performed by two 3-inch magnetron with gold and titanium targets. Sputtering were done on Si/SiO$_2$ wafer as substrate (thermally grown 100 nm SiO$_2$ on Si wafer) while the wafer were rotated at 50 rpm. Ti sputtering power set to 600 W to keep the thickness constant (under control of Ti sputtering) and Au power varied from 0 to 30 W to form homogenous layer. For formation of graded layer deposition power of Au target was ramped during anodization in the range of 0 to 30 W.

TNs layers were grown by anodization of sputtered titanium and Ti-Au sputtered layers, at 60 V in an electrolyte composed of 0.1 M ammonium fluoride (NH$_4$F) and 5 wt.% deionized H$_2$O in ethylene glycol (99 vol %) [32,33]. Two-electrode configuration was used with Ti foils as the working electrode, and platinum plate serving as the counter electrode. After anodization the samples were dipped into ethanol for 1 h. The anodization of the Ti and Ti-Au sputtered layers was conducted to completion, that is, until, full conversion of the metal films into TNs. (i.e. underlying metal layer has been used as back contact and anodization was interrupted when the current dropped due to reaching SiO$_2$ Layer. Anatase TNs layers were prepared by annealing the amorphous samples in a tube furnace at 450°C for 1 hour in air.

Samples' microstructure and morphology was investigated using an electron microscope (SEM) Hitachi FE-SEM 4800. Energy-dispersive X-ray spectroscopy (EDAX Genesis, fitted to SEM chamber) was used for the chemical analysis of sputtered and the anodic layers. The Au content determined by EDX (for both parent metal substrates and anodic oxides) and is reported by omitting O and other minor constituents, i.e. only Ti and Au concentrations are taken into account.

For open circuit photocatalytic H$_2$ evolution measurements we immersed the anodized layers in an aqueous ethanol (20 vol%) solution and illuminated the surface with 365nm LED (100

mW/cm$^2$). A gas chromatograph (GCMS-QO2010SE, SHIMADZU) with TCD detector was used to obtain the amount of generated H$_2$ for different samples.

Elemental depth profile analysis was performed according to previously published work [34], using a GD Profiler 2 from HORIBA Scientific. Plasma conditions were optimized using 650Pa for the plasma gas and a standard RF (13.56 MHz) power of 27 W. In order to reduce the sputtering rate, the RF source was pulsed at 3 kHz with a 0.25 duty cycle, giving an average power of 6.75 W. A 4 mm diameter anode was used.

**Supporting Information**

Additional characterizations and analysis supplied as supporting information. This material is available free of charge via the Internet at http://pubs.acs.org

**Acknowledgements**

The authors would like to acknowledge the ERC, the DFG, and the DFG "Engineering of Advanced Materials" cluster of excellence for financial support. The authors would also like to express their sincere thanks to Patrick Chapon (Horiba company, France) for providing GDOES measurements.

**Figure Captions**

**Figure 1.** (a) and (b) low magnification cross sectional SEM images of TiO$_2$ NTs grown on H0.9 and G$_{top}$0.6, respectively; (c) and (d) high magnification cross sectional SEM images of TiO$_2$ NTs top grown on H0.9 and G$_{top}$0.6, respectively; (e) and (f) low magnification cross sectional SEM images of TNs bottom grown on H0.9 and G$_{top}$0.6, respectively; (g) and (h) glow discharge optical emission spectroscopy depth profiles through gold contacts on TiO$_2$ NTs grown on H0.9 and G$_{top}$0.6, respectively.

**Figure 2.** (a) XRD diffractograms of pure titanium, H0.9 and G$_{top}$0.6; (b) current-time curves of different samples during anodization; (c) XRD diffractograms of TiO$_2$ NTs grown on pure titanium, H0.9 and Gtop0.6.

**Figure 3.** (a) hydrogen evolution rate of different TNs samples; (a) gold content in the sputterd Ti-Au layers; (c) gold content in TiO$_2$ NTs grown on the sputterd Ti-Au layers; (d) normalized vs. the Au loading hydrogen evolution rate of different TNs samples normalized vs. the Au loading;

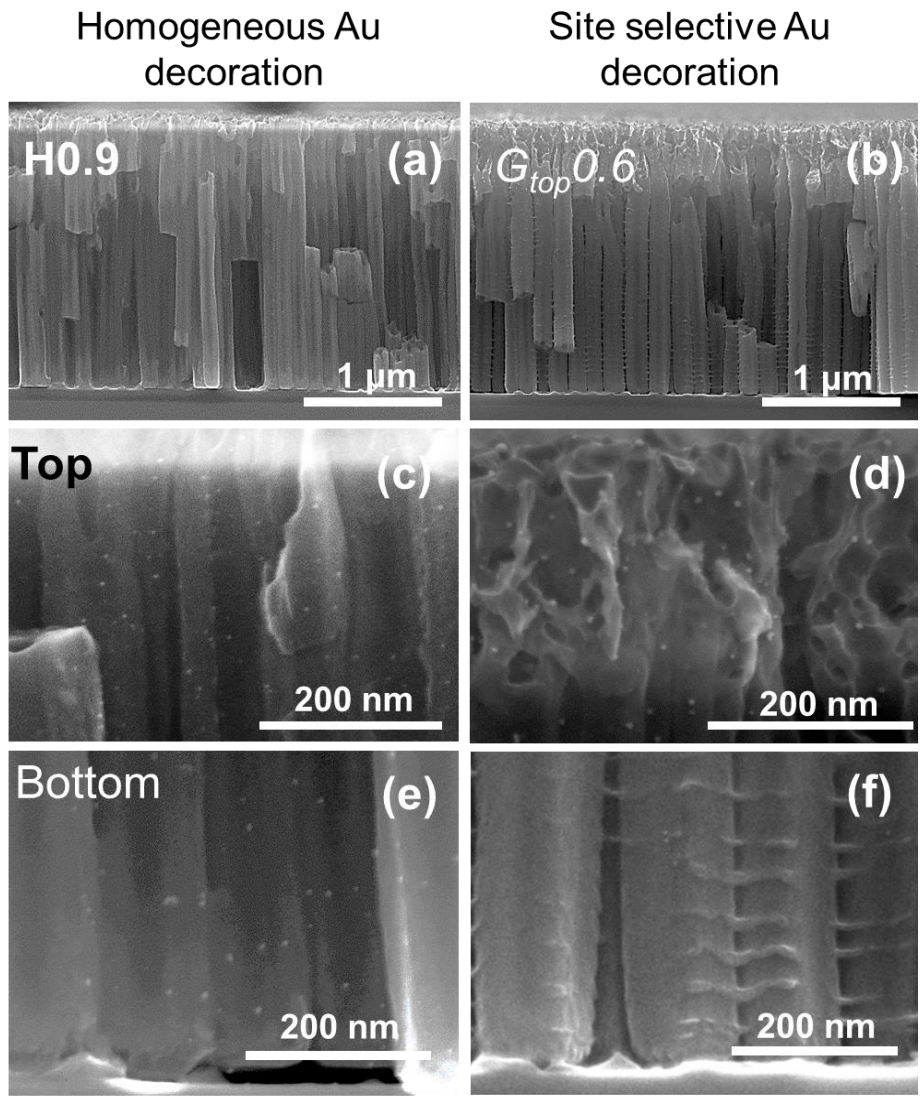
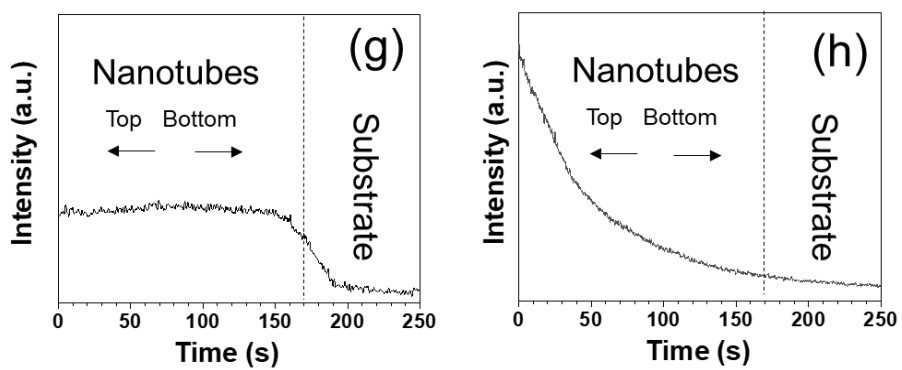

Fig. 1

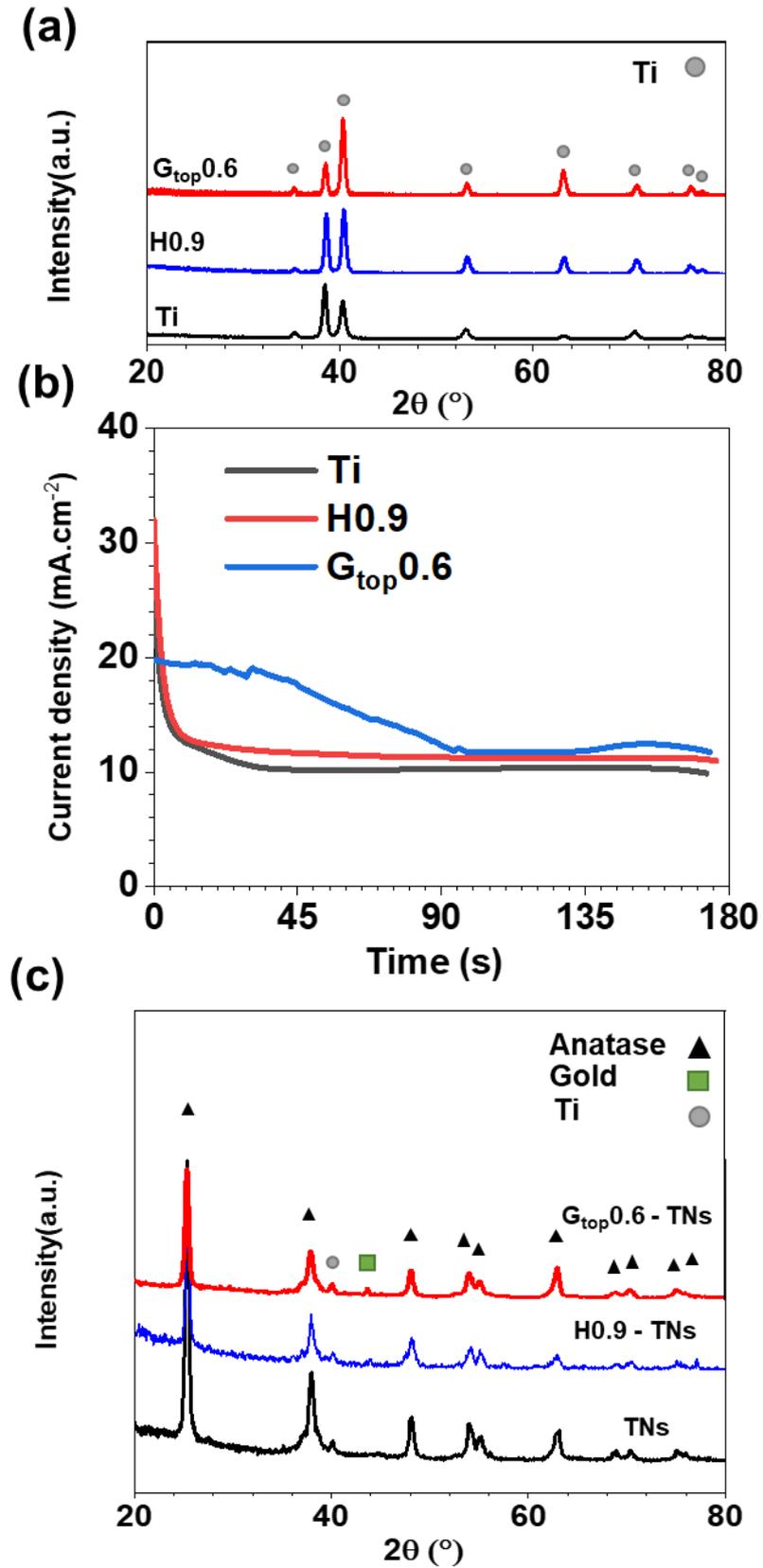

Fig. 2

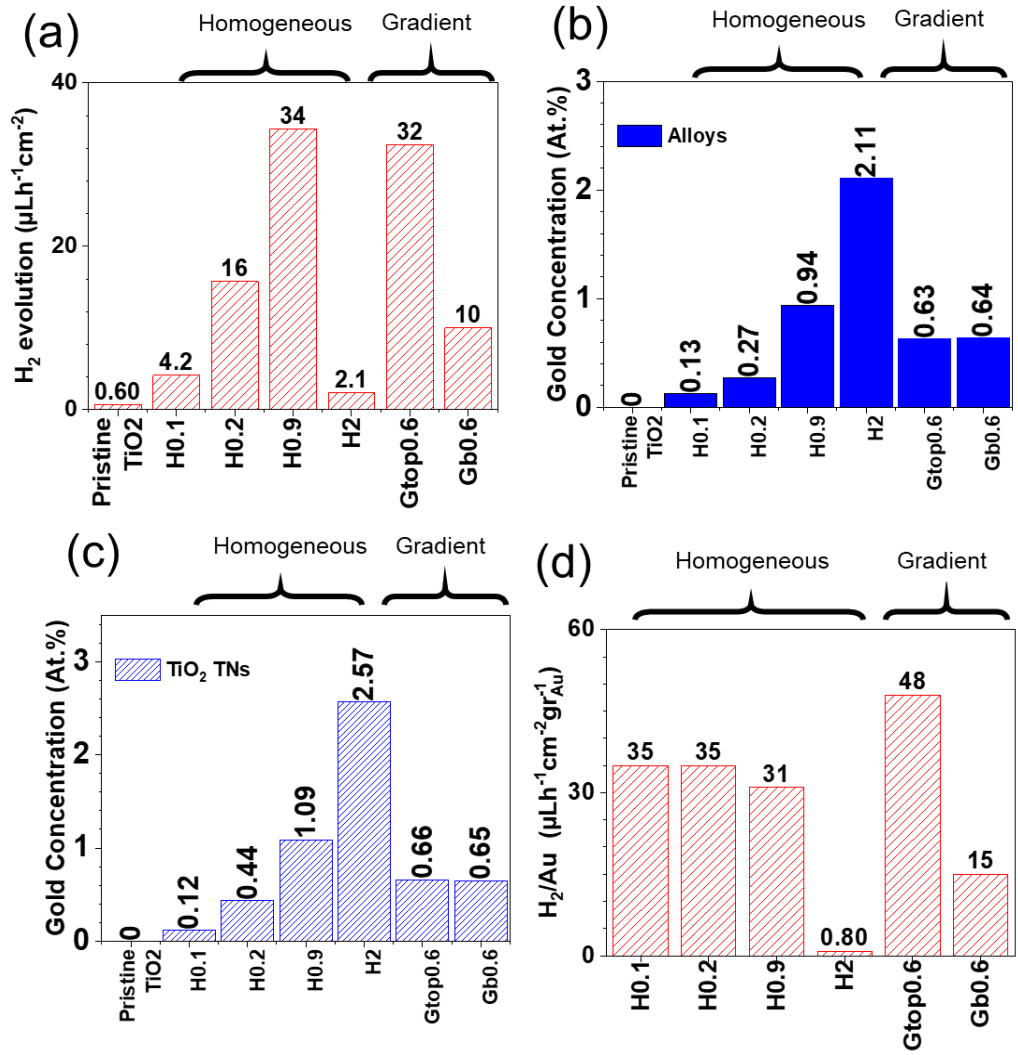

Fig. 3